\documentclass[aps,pra, 12pt,amsmath,notitlepage, tightenlines,superscriptaddress,floatfix,showpacs,showkeys,longbibliography]{revtex4-1}

\usepackage{geometry}
\usepackage{graphicx}

\usepackage{fontenc}
\usepackage{amsmath}
\usepackage{amssymb}
\usepackage{bm}
\usepackage{graphicx}
\usepackage{algorithm}
\usepackage{algpseudocode}

\usepackage{kbordermatrix}
\renewcommand{\kbldelim}{(}
\renewcommand{\kbrdelim}{)}

\usepackage{xcolor}
\definecolor{dark-red}{rgb}{0.4,0.15,0.15}
\definecolor{dark-blue}{rgb}{0.15,0.15,0.4}
\definecolor{medium-blue}{rgb}{0,0,0.5}
\usepackage{hyperref}
\usepackage[all]{hypcap}
\hypersetup{
    colorlinks, linkcolor={dark-red},
    citecolor={dark-blue}, urlcolor={medium-blue}
} 

\newcommand{\ssp}{\hspace{0.4pt}}
\newcommand{\norm}[1]{\vert #1 \vert}

\newcommand{\sH}{\mathcal{H}}
\newcommand{\sG}{\mathcal{G}}

\newcommand{\sR}{\mathcal{R}}

\newcommand{\sx}{x}
\newcommand{\sz}{z}
\newcommand{\unused}{A}

\newcommand{\supp}{\mathrm{supp}}

\newcommand{\dif}{d}
\newcommand{\omegaT}{\omega_T}
\renewcommand{\a}{a}
\renewcommand{\b}{b}
\renewcommand{\c}{c}

\newcommand{\noise}{Q}

\newcommand{\beq}{\begin{equation}}
\newcommand{\eeq}{\end{equation}}
\def\ket#1{|{#1}\rangle}
\def\bra#1{\langle{#1}|}

\begin{document}

\title{Non-commuting two-local Hamiltonians for quantum error suppression}

\date{\today}
\author{Zhang Jiang}
\email{zhang.jiang@nasa.gov}
\affiliation{Quantum Artificial Intelligence Laboratory (QuAIL), NASA Ames Research Center, Moffett Field, California 94035, USA}
\affiliation{Stinger Ghaffarian Technologies Inc., 7701 Greenbelt Rd., Suite 400, Greenbelt, MD 20770}

\author{Eleanor G. Rieffel}
\email{eleanor.rieffel@nasa.gov}
\affiliation{Quantum Artificial Intelligence Laboratory (QuAIL), NASA Ames Research Center, Moffett Field, California 94035, USA}


\begin{abstract}
Physical constraints make it challenging to implement and control 
many-body interactions. For this reason, designing quantum information 
processes with Hamiltonians consisting of only one- and two-local terms
is a worthwhile challenge. 
Enabling error suppression with two-local Hamiltonians is particularly 
challenging. 
A no-go theorem of Marvian and Lidar (Phys Rev Lett 113(26):260504, 2014)~\cite{marvian_quantum_2014}  
demonstrates that, even allowing particles with high Hilbert-space 
dimension, it is impossible to protect quantum information from single-site
errors by encoding in the ground subspace of any Hamiltonian containing 
only commuting two-local terms.
Here, we get around this no-go result by encoding
in the ground subspace of a Hamiltonian consisting of non-commuting 
two-local terms arising from the gauge operators of a subsystem code.  
Specifically, we show how to protect stored quantum information against 
single-qubit errors 
using a Hamiltonian consisting of sums of the gauge generators from 
Bacon-Shor codes (Bacon in Phys Rev A 73(1):012340, 2006)~\cite{bacon_operator_2006} 
generalized-Bacon-Shor code (Bravyi in Phys Rev A 83(1):012320, 2011)~\cite{bravyi_subsystem_2011}.  
Our results imply that non-commuting two-local Hamiltonians have 
more error-suppressing power than commuting two-local Hamiltonians. 
While far from providing full fault tolerance, this approach improves the 
robustness achievable in near-term implementable quantum storage and
adiabatic quantum computations, reducing the number of higher-order
terms required to encode commonly used adiabatic Hamiltonians such
as the Ising Hamiltonians common in adiabatic quantum optimization
and quantum annealing.

\vspace{.6cm}
\noindent \href{http://doi.org/10.1007/s11128-017-1527-9}{doi:10.1007/s11128-017-1527-9} 
\end{abstract}
\maketitle

\section{Introduction}
In order to realize the benefits of quantum computation 
\cite{NCbook,RPbook}, machines capable of robust quantum computation
must be built. An active area of research explores what resources are 
needed to achieve physically implementable robust quantum computing. 
While the ultimate goal is to achieve a fully fault-tolerant implementation,
it is also important to determine what resources enable more robust, if not 
fully fault-tolerant, quantum computations, so as to enable a richer set of 
experiments to be performed on hardware implementable in the near term.

One important question is to determine when many-body interactions are required
and when two-local interactions suffice. 
Even determining the resources required to support the robust storage of
quantum information is challenging~\cite{terhal_quantum_2015}.
Two-local interactions have been realized in all leading candidates for 
near-term quantum computation, while three-local and higher locality
interactions remain an implementation challenge. 
Within adiabatic
quantum computing, a common approach to increase robustness is
error suppression through energy gap protection (EGP)~\cite{jordan_error-correcting_2006,young_error_2013} in which
quantum information is encoded in the ground subspace of a Hamiltonian
derived from an error-detecting code
and an error-suppressing term is added to the total Hamiltonian, 
penalizing states outside that subspace. 

Unfortunately, any error-suppressing Hamiltonian constructed from a 
stabilizer code must contain interactions involving three or more qubits 
to protect against single-qubit errors.
In~\cite{marvian_quantum_2014}, Marvian and Lidar proved a stronger 
negative result, namely that energy gap protection
using an error suppressing Hamiltonian 
that is the sum of two-local commuting terms cannot suppress single-site 
errors even with particles of arbitrarily large Hilbert 
space dimension.  Their proof uses a powerful theorem due to 
Bravyi and Vyalyi~\cite{bravyi_commutative_2005}, generalized by 
Aharonov and Eldar~\cite{aharonov_complexity_2011}, which says
that ground states of commuting two-local 
Hamiltonians can have only short-range two-body entanglement.
This no-go result implies that, in the words of Marvian and Lidar,
``there is no advantage to using such codes." 
Here, we answer positively the specific question as to whether 
a Hamiltonian consisting of non-commuting two-local terms can be useful 
in combating single-qubit errors. While our claims go somewhat beyond 
single-qubit errors, we are not making broader claims about 
fault tolerance or quantum memories.

One potential way to get around the
no-go result is to try subsystem codes.  
We concentrate on stabilizer subsystem codes
\cite{bacon_operator_2006,bravyi_subsystem_2011},  
which encode information only in a subset of the qubits used by 
related stabilizer codes.
Subsystem codes offer great flexibility in designing mechanisms to
support robust quantum computation, because the effect of errors on
the unused logical qubits is not a
concern~\cite{aliferis_subsystem_2007,cross_comparative_2009}. Subsystem
codes support routines that make use of the subsystem structure to break
stabilizer terms into lower-weight terms.
However, unlike the commuting case, arising from stabilizer codes,
in which the Hamiltonians are automatically 
gapped~\cite{bravyi_commutative_2005}, there is no known general 
theorem on gaps for non-commuting Hamiltonians.  

Nevertheless, we show that indeed one can obtain single-qubit error 
suppression by encoding in the ground subspace of a Hamiltonian consisting 
of a weighted sum of non-commuting two-local terms from the gauge 
generators of a  Bacon-Shor code~\cite{bacon_operator_2006} or one of
Bravyi's generalized-Bacon-Shor codes~\cite{bravyi_subsystem_2011}.  
Both of the Bacon-Shor and the generalized-Bacon-Shor codes satisfy the 
Knill-Laflamme condition exactly, and therefore there is no ``induced 
degeneracy splitting'' that leads to the difficulties identified by 
Marvian and Lidar in the commuting case~\cite{marvian_quantum_2014}. In particular, these
codes do not come up against their generalization of the no-hiding theorem.

We explore three specific examples in detail, the $[[4,1,2]]$ Bacon-Shor 
code and the $[[6,2,2]]$ and $[[16,2,3]]$ generalized-Bacon-Shor codes.
For each of these codes, we compute the energy separation between the ground
subspace and the orthogonal subspaces. (We refer to 
the ``energy separation'' between the ground
subspace and the orthogonal subspaces as opposed to the ``gap'' so as to avoid 
confusion with the ``gap'' in an adiabatic computation, especially
when discussing error suppression in the adiabatic context.)
We provide a general technique to reduce
the dimension of the Hilbert space that needs to be considered, which eases
the calculation of the energy separation for arbitrary stabilizer 
subsystem codes.

The first code we consider provides the simplest example in which 
the construction works.
The second gives a more compact encoding and enables a reduction in the 
number of higher-order terms in encodings of commonly used adiabatic 
Hamiltonians such as the Ising Hamiltonians common in adiabatic 
quantum optimization and quantum annealing. These codes are part
of a family of codes that provide more and more compact encodings, and 
greater reductions in the number of higher-order terms, but for which the 
energy separation becomes increasingly computationally expensive to compute.
The third code provides an example that suppresses $2$-qubit errors
and provides a good example in which to see the workings of the dimension 
reduction algorithm.  
For the first two codes, we perform a numerical analysis of the open-system dynamics with the spin-boson error model, confirming exponential
suppression of single-qubit errors. We also remark briefly on the
robustness of this approach to control errors.

We expect this approach to find application in improved robustness of near-term
implementations of quantum storage, quantum annealing, and adiabatic 
quantum computing, and perhaps in quantum networks~\cite{elliott_building_2002,kimble_quantum_2008}.

\section{Brief review of quantum codes and error suppression}

The general strategy for combating local errors is to
spread the ``logical" or ``computational"
information across physical qubits. A particularly common way of
doing so is through block codes. A quantum $[[n,k]]$-block code is
a $2^k$-dimensional subspace of a $2^n$ dimensional Hilbert space, 
which enables $k$ logical qubits to be encoded in $n$ physical qubits. 
In this review section, we first give a brief review of the most common type
of quantum error-correcting codes, stabilizer codes, and then discuss 
how such codes are used to achieve error suppression. We then describe 
subsystem codes, concentrating on stabilizer subsystem codes, codes that
can be viewed as stabilizer codes in which only some of the logical 
qubits are used to encode quantum information.

\subsection{Stabilizer codes}

Consider the generalized Pauli group ${\cal P}_G$ acting on an $n$ qubit
system. Let $S$ be a subgroup of ${\cal P}_G$ generated by $r$ independent,
commuting generators. The subgroup $S$ defines a code space $C$, the 
stabilizer subspace of $S$, the joint $+1$-eigensubspace of all elements
in $S$. The dimension of $C$ is $2^k$ where $k = n-r$. 

Let $C_{{\cal P}_G}(S)$ be the centralizer of $S$, the set of elements in
${\cal P}_G$ that commute with all elements of $S$.
We define $k$ logical qubits encoded in $C$ by specifying the logical
Pauli operators in ${\cal P}_G$ that define the qubits. 
Any choice of elements outside $S$, but commuting with $S$, that satisfy the 
Pauli commutation relations works. 
There are a number of different ways of encoding, each with its own choice of
logical operators. 

The distance $d$ of a quantum error-correcting code is the minimum number of
single-qubit error by which an element of the code space can be transformed
into an orthogonal element of the code space. The weight of a Pauli error 
$e\in {\cal P}_G$ is the number of qubits on which a non-identity Pauli 
transformation acts. An $[[n,k,d]]$-quantum code can detect up to weight
$d-1$ errors, and can correct errors of weight up to $t$ satisfying $2t < d$.

We now give the $[[4,2,2]]$-stabilizer code as an example since it will set
us up well for discussing the $[[4,1,2]]$-subsystem code. 
Consider a $4$-qubit system consisting of qubits $q_{1,1}$, $q_{1,2}$, 
$q_{2,1}$, and $q_{2,2}$. 
Let $C$ be the joint $+1$-eigensubspace of the two stabilizer generators
\begin{align*}
\begin{split}
 &S^X = X_{1,1}X_{1,2}X_{2,1}X_{2,2}\\
 &S^Z = Z_{1,1}Z_{2,1}Z_{1,2}Z_{2,2}.
\end{split}
\end{align*}

We may define logical operators that define two encoded qubits as:
\begin{align*}
\begin{split}
 &\tilde X_{L1} = X_{1,1}X_{2,1}\;,\quad \tilde Z_{L1} = Z_{1,1}Z_{1,2} \\
 &\tilde X_{L2} = X_{1,1}X_{1,2}\;,\quad \tilde Z_{L2} = Z_{1,1}Z_{2,1}.
\end{split}
\end{align*}
The reader may check that $X_{L1}$, $Z_{L1}$, $X_{L2}$, and $Z_{L2}$ are in
$C_{{\cal P}_G}(S)$, satisfy the Pauli commutation relations, and that the 
code distance is $2$. A useful property of stabilizer codes is that any 
operator that is the product of a logical operator and
stabilizers behaves on the code space in exactly the same way as the
logical operator behaves on the code space since the stabilizers
commute with logical operators.

\subsection{Error suppression with stabilizer codes}

Let $H_0(t)$ be a problem Hamiltonian acting on an $n$-qubit system. 
Given an $[[n,k,d]]$-stabilizer code, the Hamiltonian
$$H(t) = H_L(t) + E_P H_\supp$$
implements error suppression, where $E_P$ is the penalty weight, 
$H_\supp$ is the negative of the sum of $M$ stabilizers,
and $H_L(t)$ 
is obtained from $H_0(t)$ by replacing all operators with their logical
counterpart. Any error of weight less than $d$ will take the state out of the 
$+1$-eigensubspace of at least one stabilizer, and thus receives a
penalty from $H_\supp$.
As an example, the problem Hamiltonian
\beq
H_0(t) = a(t)\sum_{i=1}^n X_i + b(t)\left(\sum_{i=1}^n h_iZ_i
+ J_{ij}Z_iZ_j\right)
\eeq
becomes the encoded Hamiltonian
\beq
H(t) = a(t)\sum_{i=1}^n \tilde X_i + b(t)\left(\sum_{i=1}^n h_i\tilde Z_i
+ J_{ij}\tilde Z_i\tilde Z_j\right) + E_P H_\supp.
\eeq
By construction, the logical Hamiltonian $H_L(t)$ commutes with the 
error suppression term $H_\supp$, so in the error-free case, the logical 
dynamics of the 
code space under $H(t)$ is that of the original space under $H(t)$.

The Hamiltonian $H_\supp$ is 4-local for the $[[4,1,2]]$ code,
since its stabilizers act on all four physical qubits. The energy
separation of $H_\supp$ is $E_P$, because the sum of the two commuting
stabilizers has eigenvalues $-1,0,1$.
In the next subsection, we will see how the 4-local stabilizers can be
broken into 2-local gauge operators by encoding the quantum information in a
subsystem instead of a subspace.

\subsection{Subsystem codes}

Subsystem codes decompose the Hilbert space into 
$\sH = \sH_L\otimes \sH_G$, where $\sH_L$ is the subsystem of
logical qubits that store information, and $\sH_G$ is the subsystem
of the gauge qubits. We concentrate on stabilizer subsystem 
codes~\cite{poulin_stabilizer_2005,bacon_operator_2006}, 
a generalization of stabilizer codes, particularly  
2D generalized-Bacon-Shor codes.
Stabilizer subsystem codes
can be viewed as stabilizer codes in which only some of the logical
qubits are used to encode quantum information.  

A 2D generalized-Bacon-Shor code is specified by a 2D array, or
matrix, $M$ with entries in $\mathbb Z_2$.
For each $1$ in the matrix, we have a physical qubit.
The matrix defines a $[[n, k, d]]$ code where $n$ is the number of nonzero
matrix elements in $M$, $k$ is the rank of $M$, and $d$ is the minimum
distance of the two classical codes defined by the rows and columns of $M$,
respectively.

The gauge group $\cal G$ is generated by $XX$ operators for every pair 
of qubits
in the rows of $M$ and $ZZ$ operators for every pair of qubits in the
columns of $M$. There is redundancy in this generating set. A commonly used
non-redundant generating set is the set of generators
corresponding to nearest neighbor pairs.
The stabilizer subgroup of the code is the center of $\cal G$,
the set of elements of $\cal G$ that commute with
all elements of $\cal G$. The reader can check that the stabilizer subgroup
consists of all products of $X$ operators corresponding
to linearly dependent sets of rows and all products of $Z$ operators
corresponding to linearly dependent sets of columns.

The logical operators are chosen
to be elements of $\cal C(G)\setminus G$, elements of the centralizer of
the gauge group that are not in the gauge group. Just as in the
stabilizer code case, there is freedom in the choice of logical operators. 
We define $k$ logical qubits by specifying logical
Pauli operators in $\cal C(G)\setminus G$ that satisfy the 
Pauli commutation relations. 

It is useful to define auxiliary qubits, subspaces of the gauge subspace
that are not in the stabilizer subspace. These qubits will not be used to 
encode computational information, but will be useful in analyzing
subsystem codes.

\emph{Example:} The $[[4,1,2]]$-stabilizer subsystem code.
\begin{figure} 
   \centering
   \includegraphics[width=0.4\textwidth]{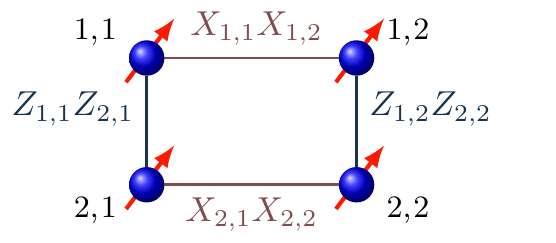}
   \caption[Bacon-Shor code]{The gauge generators of the Bacon-Shor 
$[[4,1,2]]$-code consist of two-qubit operators of type $XX$ ($ZZ$) 
that couple qubits in a row (column).}
   \label{fig:412_code}
\end{figure}

The most familiar subsystem code is the Bacon-Shor $[[9,1,3]]$-code, 
the smallest Bacon-Shor code that corrects single-qubit errors. 
For error suppression, codes that only {\em detect} errors
can be used, allowing us to consider the smallest error-detecting
Bacon-Shor code, the $[[4,1,2]]$-code, which detects single-qubit 
errors. This code has been used before, 
by Brell \emph{et al.}~\cite{brell_toric_2011}, as a gadget to obtain 
the Hamiltonians for the toric code and Kitaev's quantum double models 
as the low-energy limits of two-body Hamiltonians. 

The gauge group for the $[[4,1,2]]$-code is 
\begin{align}
 \sG &= \big\langle G^X_1,\, G^X_2,\,G^Z_1,\,G^Z_2\big\rangle\;,
\end{align}
which is generated by the gauge generators
\begin{align}
\begin{split}
 &G^X_1 = X_{1,1}X_{1,2}\,,\quad G^X_2= X_{2,1}X_{2,2}\,,\\
 &G^Z_1 = Z_{1,1}Z_{2,1}\,,\quad G^Z_2 = Z_{1,2}Z_{2,2}\,,
\end{split}
\end{align}
shown in Figure~\ref{fig:412_code}.  
The stabilizer subgroup is generated by
\begin{align}
\begin{split}
 &S^X = X_{1,1}X_{1,2}X_{2,1}X_{2,2}\;,\\
 &S^Z = Z_{1,1}Z_{2,1}Z_{1,2}Z_{2,2}\;.
\end{split}
\end{align}
The reader will recognize these stabilizers as those defining
the $[[4,2,2]]$-stabilizer code. 
The $[[4,1,2]]$-subsystem code encodes information in only one of 
the logical qubits of the $[[4,2,2]]$-stabilizer code. 
One choice of logical operators is
\begin{align}
\begin{split}
 &X_L = X_{1,1}X_{2,1}\;,\quad Z_L = Z_{1,1}Z_{1,2}\;.
\end{split}
\end{align}
A convenient choice of auxiliary operators is 
\begin{align}
\begin{split}
 &X_\unused = G^X_1 = X_{1,1}X_{1,2} \;,\quad Z_\unused = G^Z_1 = Z_{1,1}Z_{2,1} \;.
\end{split}
\end{align}
These operators would have defined a logical qubit in a stabilizer code,
but in the subsystem code this qubit is not used to encode computational
information.

Examples of larger subsystem codes with more complex structure will be
given in later sections.

\section{Two-local Hamiltonians for error suppression}

Given a Hamiltonian $H(t)$ we wish to encode to suppress errors and a specific 
subsystem code, we create an encoded Hamiltonian 
$H_E(t) = H_L(t) + H_\supp$, 
where $H_L(t)$ is obtained from $H$ by replacing each operator in $H$ with the
corresponding operator for the code and where $H_\supp$ is a weighted sum
of the gauge operators. Because the logical operators commute with the
gauge group, the evolution of the encoded subspace under $H_E(t)$ is that
of $H(t)$ acting on an unencoded system, so the dynamics are correct. 

Unlike the stabilizer code case, in which all the terms of $H_\supp$ commute
and therefore the Hamiltonian is known to be gapped, we must check 
that $H_\supp$ has an energy separation between the ground subspace and 
orthogonal subspaces. Because the gauge group of any stabilizer subsystem
code is generated by two-local operators, the suppression term $H_\supp$
is always two local. As we will see, whether $H_L$ is two-local depends 
on properties of $H$ and of the subsystem code.

\subsection{Suppression with the Bacon-Shor $[[4,1,2]]$-code} 

Consider an error-suppressing Hamiltonian that is a weighted sum of the 
gauge generators, where the weights $\lambda_i$ and $\eta_i$ are real:  
\begin{align}\label{eq:supp_H_412}
\begin{split}
 H_\mathrm{supp} &= -\lambda_1 G^X_1-\eta_1 G^Z_1 - \lambda_2 G^X_2 - \eta_2 G^Z_2\\
 & = -\lambda_1 X_\unused-\eta_1 Z_\unused - \lambda_2 S^X  X_\unused - \eta_2 S^Z Z_\unused\\
 & = -(\lambda_1 + \lambda_2 S^X) X_\unused -(\eta_1 + \eta_2 S^Z) Z_\unused\;,
\end{split}
\end{align}
where the last line uses the relation
$G^X_2 = S^X X_\unused$ and $G^Z_2 = S^Z Z_\unused$.  
This Hamiltonian, being a linear combination 
of the gauge generators, commutes with the stabilizers and logical operators.  
It takes a block diagonal form in the eigenbasis of the stabilizers. 
The energy separation of $H_\mathrm{supp}$ is a function of
$\lambda_i$ and $\eta_i$, which can be calculated exactly in this case.
For fixed values of the stabilizers, we have 
\begin{align}\label{eq:supp_H_412_xz}
H_\mathrm{supp}^{(x,\, z)}
= -(\lambda_1 + \lambda_2 x) X_A -(\eta_1 + \eta_2 z) Z_A\;,
\end{align}
where $\sx,\,\sz = \pm 1$ are the eigenvalues of $S^X$ and $S^Z$, respectively.
Any detectable error inevitably changes the value of
the stabilizers, and is suppressed by the energy separation
of $H_\mathrm{supp}$.
The eigenvalues of this Hamiltonian are
\begin{align}
 E_\mathrm{supp}^{(\sx,\, \sz)} = \pm \sqrt {(\lambda_1+\sx\lambda_2)^2 + (\eta_1+\sz\eta_2)^2} \;.
\end{align}
For the case where $\lambda_1,\lambda_2 > 0$ and 
$\eta_1, \eta_2 > 0$, the ground subspace
of $H_\mathrm{supp}^{(\sx^\star\!,\,\sz^\star)}$ with $\sx^\star=\sz^\star=1$ has energy 
strictly smaller than that of $E_\mathrm{supp}^{(\sx,\, \sz)}$ with
$(\sx,\, \sz) \ne (1,1)$.  
Any single-qubit error on logical information encoded in the ground
state of $H_\mathrm{supp}^{(\sx^\star\!,\,\sz^\star)}$ is suppressed, 
because it raises the energy
by changing the value of $\sx^\star$ or $\sz^\star$.  
For $\lambda_1 = \lambda_2 = \eta_1 = \eta_2 = 1$, the eigenvalues are
$\pm\sqrt{8}, \pm 2, 0$, so the gap is $2(\sqrt{2} - 1)$.
Thus, unlike in the stabilizer code case, a two-local Hamiltonian can
suppress arbitrary single-qubit errors.

\subsection{Suppression with the Bacon-Shor $[[6,2,2]]$-code} 
We now turn to Bravyi's 
$[[6,2,2]]$ generalized-Bacon-Shor code~\cite{bravyi_subsystem_2011}. 
It has the advantage of being more
compact, encoding two logical qubits in six physical qubits, and has the 
further advantage, as we will see, that certain two-qubit logical operators 
can be implemented using operations involving only two physical qubits.

\begin{figure} 
   \centering
   \includegraphics[width=0.35\textwidth]{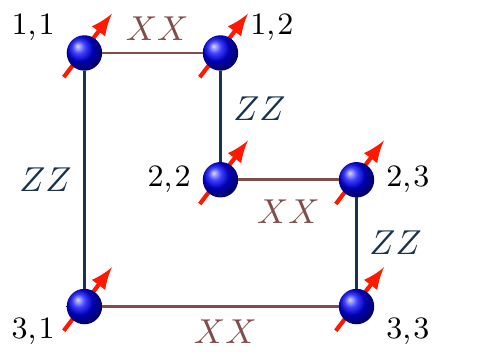}
   \caption[Bravyi's subsystem code]{The gauge generators of 
Bravyi's $[[6,2,2]]$-code formed using $6$ physical qubits 
placed within a $3\times 3$ lattice. The gauge generators consist 
two-qubit operators of type $XX$ ($ZZ$) coupling qubits in a row (column).}
   \label{fig:622_code}
\end{figure}

The following matrix defines Bravyi's $[[6,2,2]]$-code 
\begin{align}
 M = 
 \begin{pmatrix}
  1 & 1 & 0\\
  0 & 1 & 1\\
  1 & 0 & 1
 \end{pmatrix}\;. 
\end{align}
The gauge group of Bravyi's code is
\begin{align}
 \sG &= \big\langle G^X_1,\, G^X_2,\,G^X_3,\,G^Z_1,\,G^Z_2,\,G^Z_3\big\rangle\;,
\end{align}
generated by the gauge generators
as shown in Figure~\ref{fig:622_code}:
\begin{align}
\begin{gathered}
 G^X_1 = X_{1,1}X_{1,2}\,,\quad G^X_2= X_{2,2}X_{2,3}\,,\quad G^X_3= X_{3,1}X_{3,3}\,,\\
 G^Z_1 = Z_{1,1}Z_{3,1}\,,\quad G^Z_2 = Z_{1,2}Z_{2,2}\,,\quad G^Z_3 = Z_{2,3}Z_{3,3}\;.
\end{gathered}
\end{align}
The center of $\sG$ is generated by the stabilizers
\begin{align}
\begin{split}
 S^X &=  G^X_1 G^X_2 G^X_3\;, \quad
 S^Z = G^Z_1 G^Z_2 G^Z_3\;. 
 \end{split}
\end{align}
One choice for the logical operators is
\begin{align}
\begin{split}
 &X_{L1} = X_{2,3}X_{3,3}\;,\quad Z_{L1} =  Z_{3,1}Z_{3,3}\,,\\
 &X_{L2}  = X_{1,2}X_{2,2}\;, \quad  Z_{L2} = Z_{1,1}Z_{1,2}\,.
\end{split}
\end{align}

The auxiliary logical operators can be chosen to be the gauge elements
\begin{align}
\begin{split}
 &X_{\unused 1} = X_{3,1}X_{3,3} = G^X_3\;,\quad Z_{\unused 1} = Z_{2,3}Z_{3,3} = G^Z_3\,,\\
 &X_{\unused 2} = X_{1,1}X_{1,2} = G^X_1\;, \quad  Z_{\unused 2} = Z_{1,2}Z_{2,2} = G^Z_2\,.
\end{split}
\end{align}
The original gauge generators can be expressed using the stabilizers and the unused logical operators
\begin{align}
\begin{gathered}
 G^X_1 = X_{\unused 2}\,,\quad G^X_2 = S^X X_{\unused 1} X_{\unused 2}\,,\quad G^X_3 = X_{\unused 1}\,,\\
 G^Z_1 = S^Z Z_{\unused 1} Z_{\unused 2}\,,\quad G^Z_2 = Z_{\unused 2}\,,\quad G^Z_3 = Z_{\unused 1}\;.
 \end{gathered}
\end{align}


Consider an error-suppressing Hamiltonian that is a weighted sum of the 
gauge generators 
\begin{align}\label{eq:supp_H}
\begin{split}
 H_\mathrm{supp} &= \mathord - \lambda_1 G^X_3 - \lambda_2 G^X_1 -  \lambda_1' G^Z_3 - \lambda_2' G^Z_2 - \eta\, G^X_2   - \eta'\, G^Z_1\;,
\end{split}
\end{align}
where all coefficients are positive real numbers. 
For fixed values of the stabilizers $S^X = x$ and $S^Z = z$, the 
Hamiltonian~(\ref{eq:supp_H}) can be written in terms of the auxiliary 
logical operators
\begin{align}\label{eq:supp_H_s}
\begin{split}
 H_\mathrm{supp}^{(\sx,\ssp\sz)}  &= \mathord - \lambda_1 X_{\unused 1} - \lambda_2 X_{\unused 2} -  \lambda_1' Z_{\unused 1} - \lambda_2' Z_{\unused 2} - \eta\ssp \sx X_{\unused 1} X_{\unused 2} - \eta' \sz Z_{\unused 1} Z_{\unused 2} \;.
\end{split}
\end{align}
We now demonstrate that, as for the $[[4,1,2]]$ code, 
the energy separation of the above Hamiltonian for the $[[6,2,2]]$ code is
nonzero for typical choices of the weights.
To suppress errors detectable by
the code, the ground states of $H_\mathrm{supp}$ should also be eigenstates of
the stabilizers with 
fixed eigenvalues $\sx=\sx^\star$ and $\sz=\sz^\star$. 
Consequently, the Hamiltonian $H_\mathrm{supp}^{(\sx,\ssp\sz)}$ must 
depend on the values of $\sx$ or $\sz$, so we must have 
$\eta\neq 0$ and $\eta'\neq 0$.  In general, to achieve error suppression,
none of the coefficients in 
Eq.~(\ref{eq:supp_H_s}) can be zero. 
For example, we must have $\lambda_1\neq 0$; 
otherwise, the eigenvalues of $H_\mathrm{supp}^{(\sx,\ssp\sz)}$ would
not depend on the value of $\sx$.  
We can show this by simply flipping the sign of the term with coefficient $-\eta x$ by applying the unitary transformation 
$Z_{\unused 1} H_\mathrm{supp}^{(\sx,\ssp\sz)} Z_{\unused 1}^\dagger$, 
without changing anything else.  

The general case can be solved directly. A particular choice of the error-suppressing Hamiltonian that can be solved
easily is \begin{align}
\begin{split}
 H_\mathrm{supp}^{(\sx,\ssp\sz)}  &= \mathord - \lambda\big(X_{\unused 1} + X_{\unused 2} +  Z_{\unused 1} + Z_{\unused 2}\big) - \eta\big(\sx X_{\unused 1} X_{\unused 2} + \sz Z_{\unused 1} Z_{\unused 2}\big) \;,
\end{split}
\end{align}
where $\lambda, \eta >0$. 
In the Bell basis of the two auxiliary logical qubits, we have
\begin{align}\label{eq:gauge_H}
 H_\mathrm{supp}^{(s_\pm)} = \mathord - 2\,
 \begin{pmatrix}
  \eta s_+ & \lambda & \lambda & 0\\
  \lambda &  \eta s_-  & 0 & 0\\
  \lambda & 0 &- \eta s_- & 0\\
  0 & 0 & 0 &- \eta s_+\\
 \end{pmatrix}\;,
\end{align}
where $s_\pm = (\sx \pm \sz)/2 \in \{-1,\,0,\,1\}$, and the singlet state is always decoupled from the other states.  
The eigenvalues of this matrix are $\big\{\mathord - \eta s_+ \pm \sqrt{8\lambda^2 + \eta^2},\, 2\eta s_+,\, 0\big\}$, for $s_+=\pm 1$ and $s_-=0$, 
and $\big\{\mathord\pm 2\sqrt{2 \lambda^2 + \eta^2},\, 0,\, 0\big\}$, 
for $s_+=0$ and $s_-=\pm 1$. 
Thus, the ground subspace for the case $\sx=\sz=1$ has the lowest 
energy $-\big(\eta + \sqrt{8\lambda^2 + \eta^2}\,\big)$, and the two 
ground states for $\sx= - \sz = \pm 1$ have 
the second-lowest energy $-2\sqrt{2\lambda^2 + \eta^2}$.  The gap is 
$\eta + \sqrt{8\lambda^2 + \eta^2} - 2\sqrt{2\lambda^2 + \eta^2}$. The
gap goes to $\eta$ for $\lambda \rightarrow \infty$, and is zero for 
$\lambda \rightarrow 0$. 
It equals to $4 -2 \sqrt 3\simeq 0.536$ for $\lambda =\eta = 1$. 

\subsection{Reducing number of higher-order terms in logical Hamiltonians}

While the error suppression term coming from these subsystem code constructions
is two local, the logical Hamiltonian may contain higher-order terms.
For example, an Ising Hamiltonian has $ZZ$ terms. The corresponding
logical Ising Hamiltonian, implemented using two two-local 
logical $Z$ operators, would in general be four local.

In addition to compactness, another
advantage of the $[[6,2,2]]$-code over the $[[4,1,2]]$-code
is that the logical operators acting on two qubits encoded together
are two-local. Because the stabilizer terms commute with the encoded logical
Hamiltonian, $X_{L1}X_{L2}$ can be implemented using $X_{1,1}X_{3,1}$
and $Z_{L1}Z_{L2}$ can be implemented using $Z_{2,2}Z_{2,3}$ since
\begin{align}
\begin{split}
 &X_{L1} X_{L2} = S^X X_{1,1}X_{3,1}\;,\quad 
 Z_{L1} Z_{L2} = S^Z Z_{2,2}Z_{2,3}\;,
\end{split}
\end{align}
so certain two-qubit logical operators, those acting on two logical 
qubits encoded together, can be implemented with two-local interactions only. 
When multiple qubits are encoded using multiple copies of the code, two-qubit
logical operators acting on logical qubits encoded by different copies
will still need four-body interactions. 
Nevertheless, this code has an advantage over the $[[4,1,2]]$-code 
in which every logical qubit is encoded separately so one needs four-body 
interactions to implement every two-qubit logical operation.

These codes are the first two examples of codes in a family of
$[[2k+2, k, 2]]$-codes specified by $(k+1)\times(k+1)$-matrices $M$
with row $i$ containing $1\,$s only in positions $i$ and $i+1$ (where
the $k+1$st row contains $1\,$s in positions $k+1$ and $1$). Unfortunately,
the energy separation becomes more and more 
computationally intensive to compute as $k$ increases.

\subsection{Generalized construction}

We have seen that it is useful to write the error-suppressing
Hamiltonian in terms of auxiliary operators and stabilizers. 
Doing so removes the degeneracy in $H_\supp$ and reduces 
the size of the Hilbert space under consideration, enabling 
numerical calculation of the energy separation between the code subspace 
and orthogonal subspaces. Such a reduction is always possible for 
generalized-Bacon-Shor codes; in Appendix~\ref{sec:auxiliary} we give an 
algorithm that provides a systematic way of finding auxiliary operators 
satisfying the standard commutation relations that, together with the 
stabilizer operators, generate the gauge group. 

We made use of this algorithm to compute the energy separation for a
$[[16,2,3]]$ generalized-Bacon-Shor code (see Appendix~\ref{sec:auxiliary} for more details). Again, we may encode information 
in the ground subspace with all stabilizers taking value $+1$.
Then, for $\lambda = 1$, when the error suppression term is 
precisely the (unweighted) sum of all gauge generators,
the ground subspace energy is $-13.83$ and the energy separation
between the code space and the orthogonal subspaces is $0.33$.

We can apply this construction to generalized-Bacon-Shor codes that 
encode a larger number of qubits at once, thus increasing the proportion
of logical operators that can be implemented using only two-local
interactions. However, for these larger codes, computing the energy 
separation becomes computationally prohibitive.  As was shown
in~\cite{dorier_quantum_2005} using exact diagonalizations, Green’s function
Monte Carlo simulations, and high-order perturbation theory, the energy
separation of a Bacon-Shor code (or the quantum compass model) vanishes
exponentially in the size of the code. An open question is whether it is
possible to obtain a nonzero asymptotic separation for some family of
generalized-Bacon-Shor codes.

\subsection{Robustness to implementation errors} 
Our analysis shows that effective error suppression can be obtained
using a variety of different weighted sums of the gauge generators
as the $H_\supp$ term. For this reason, this approach is highly robust
to control errors that result in imprecise implementation of the
weights of these terms.
Furthermore, when the weight of any implementation error in the logical
Hamiltonian is less than the code distance, these errors are suppressed
by $H_\supp$. 
This robustness does not yield fault tolerance, but does mean that
this error suppression approach can be useful in near-term
implementations of quantum computational devices and quantum storage.

\section{Numerical analysis of effectiveness of error suppression with 
spin-boson noise model}

We perform a numerical analysis of the effectiveness 
these codes in suppressing errors due to qubits coupling to individual
baths. Specifically, we show that decoherence effects on qubits coupled
to bosonic baths with Ohmic spectra are exponentially suppressed when 
the energy separation of $H_\mathrm{supp}$ becomes larger than $k_\mathrm{B}T$.
A physical interpretation for that condition is that the bath modes, which 
resonate with the transition frequencies of the system to higher-energy 
states, are close to the vacuum state at the temperature $T$ 
and thus are not capable of driving such transitions. 

\subsection{Spin-boson noise model}

The spin-boson error model~\cite{leggett_dynamics_1987} we use
generalizes the one used 
in~\cite{pudenz_error-corrected_2014} in which individual system qubits 
are coupled to independent baths, but via all Pauli operators not just $Z$
operators,
\begin{align}\label{eq:noise_analysis_hamiltonian}
\begin{split}
 H &=  H_\supp + H_B + \sum_{k=1}^n \Big(X_k\otimes B_k^X+  Y_k\otimes B_k^Y + Z_k\otimes B_k^Z\Big)\;,
 \end{split}
\end{align}
where $H_B$ is the bath Hamiltonian, $B_k^{X}$ is the bath operator 
coupled to the $X$ operator of the $k$th qubit, and similarly for 
$B_k^{Y}$ and $B_k^{Z}$ (see the Appendix~\ref{sec:noise_model} for more details). 

We assume that the bath spectral functions $C_\mathrm{bath}(\omega)$ 
are the same for all physical qubits. Its Fourier transformation takes 
the following form for the Ohmic case,
\begin{align}
 \widetilde C_\mathrm{bath}(\omega)  =  \frac{2 \pi \hbar^2 \chi \omega\ssp e^{-\norm{\omega}/\omega_c}}{1- e^{-\hbar\omega/k_\mathrm{B} T }}\;,
\end{align}
where $T$ is the temperature, $\omega_c$ is the cutoff frequency, 
and $\chi$ is a dimensionless parameter proportional to the product of the system-bath 
coupling strength and the bath spectral density. 
For $\omega < 0$ (a transition that raises the energy of the system) 
and $\norm{\omega}$ sufficiently large, the term in the denominator 
becomes most salient, yielding exponential suppression of decoherence.
In our analysis, we use
the parameters from~\cite{pudenz_error-corrected_2014}: $\chi = 3.18 \times 10^{-4}$, $\omega_c=8\pi\times 10^9\, \mathrm{rad/s}$, and $k_\mathrm{B} T/\hbar = 2.2\times 10^9\, \mathrm{rad/s}$ (at $17\,\mathrm{mK}$). 

\subsection{Numerical analysis of error suppression} 

After going to the interaction picture of $H_\mathrm{supp}$, we
simulate the open-system dynamics of the qubits using the Markov approximation.
The bath correlation function $\widetilde C_\mathrm{bath}(\omega)$ determines
the transition rate between two states in two different energy eigenspaces of
$H_\mathrm{supp}$. A transition is suppressed when the energy increase is
significantly greater than $k_\mathrm{B} T/\hbar$.

\begin{figure}
   \includegraphics[width=0.48\columnwidth]{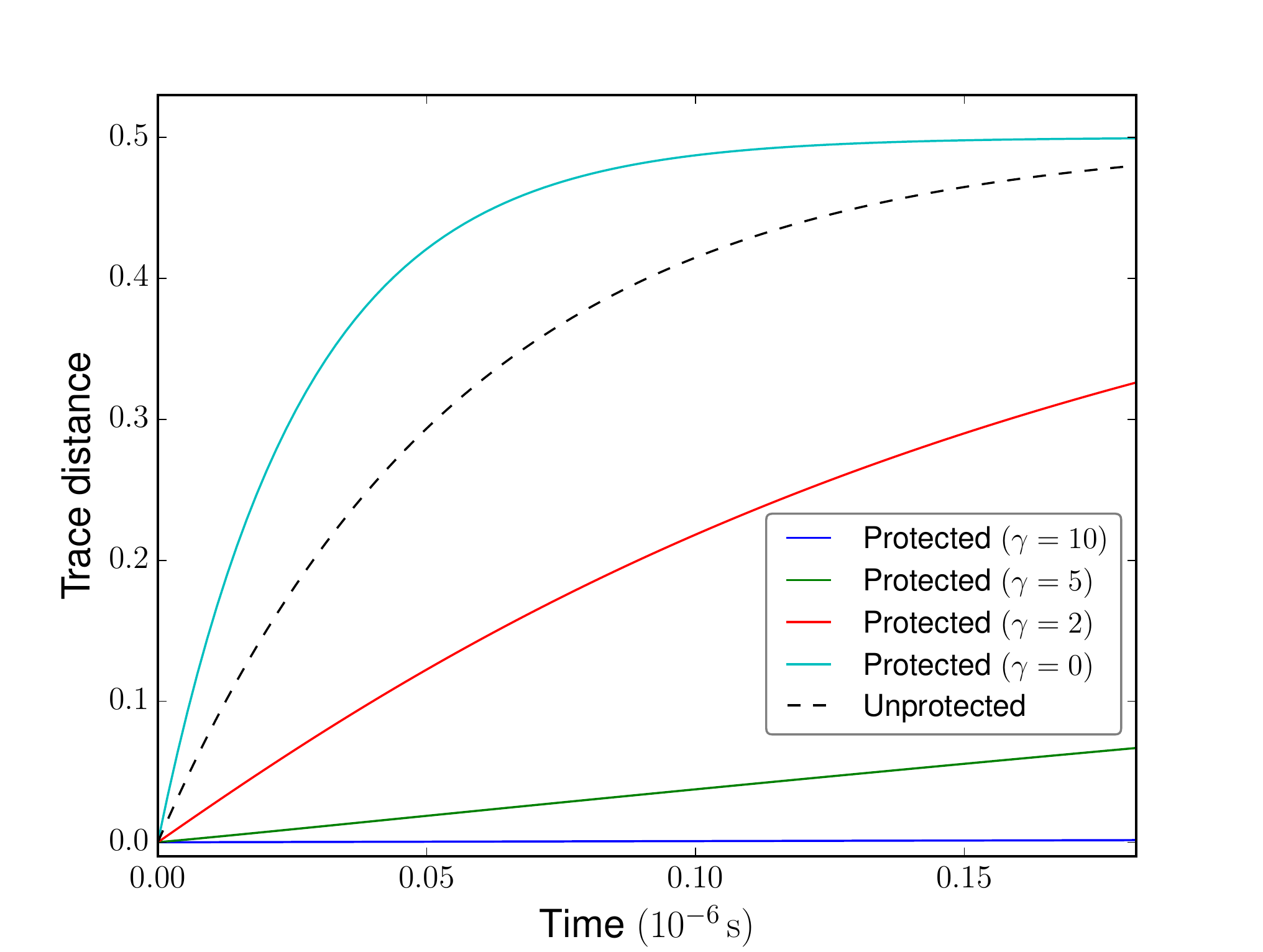}
   \includegraphics[width=0.48\columnwidth]{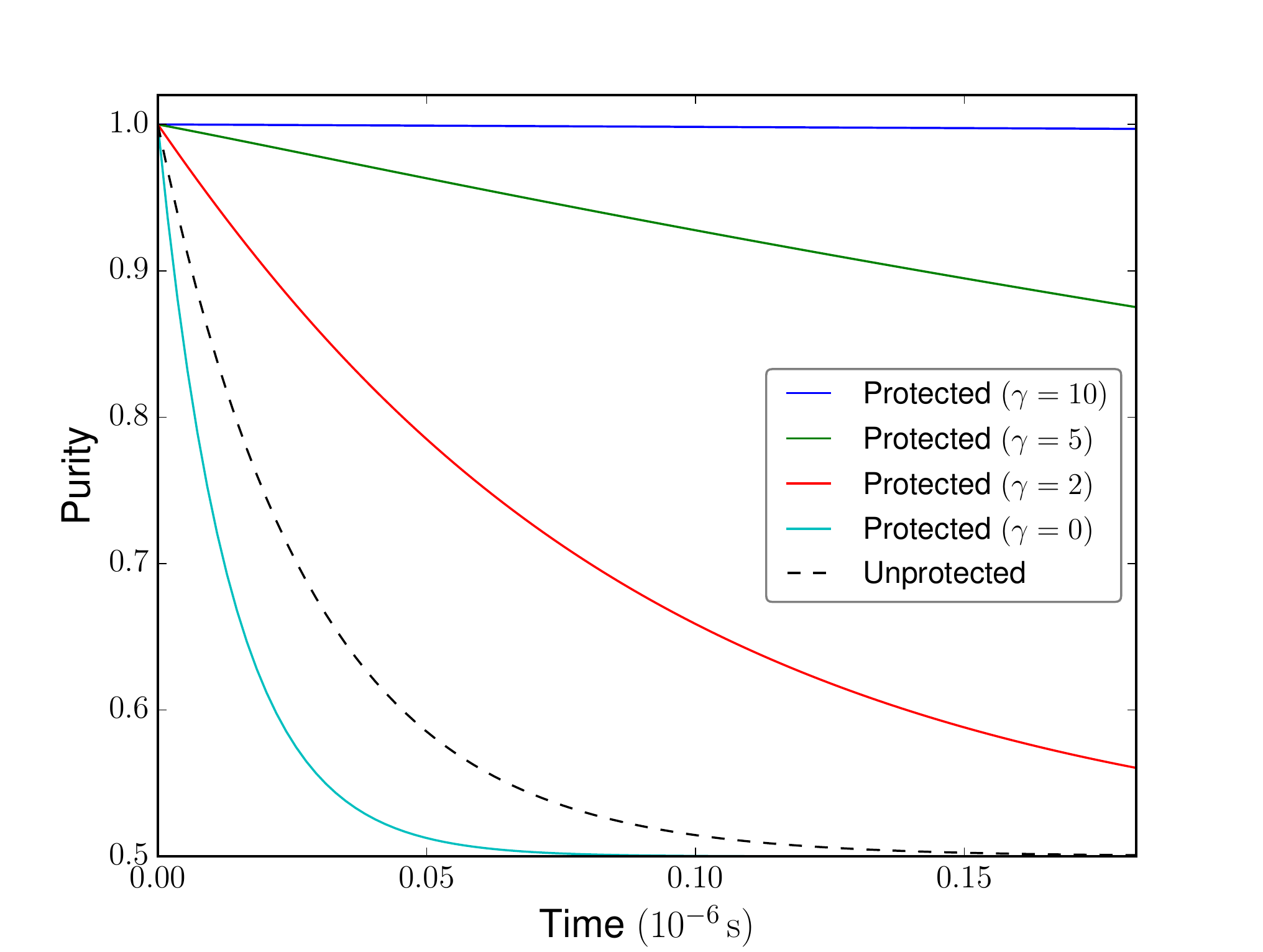}
   \caption{Simulations of the open-system dynamics of a single logical qubit,
encoded in the $[[4,1,2]]$ subsystem code, 
under single-qubit noise for different penalty weights $\gamma=\lambda/k_\mathrm{B}T$. 
The initial state of the logical
qubit is set to be the state $\ket{+}_L$, the $+1$ eigenstate of $X_L$.  
(\textit{Left}) Trace distance between the initial and evolved state.  
(\textit{Right}) Purity of the evolved state.}
   \label{fig:simulate_412}
\end{figure}

For various codes, we consider an error-suppressing Hamiltonian proportional
to the sum over a generating set $\sR$ for $\cal G$, 
$H_\supp = -\lambda\sum_{G\ssp \in\ssp \sR} G$. 
In Figure~\ref{fig:simulate_412} we show results for the $[[4,1,2]]$ 
subsystem codes at various implementable values of $\gamma = \lambda/ k_\mathrm{B} T$.
Error suppression is helpful once
$\gamma$ is larger than a threshold value of $\simeq 0.6$. 
Below that value, the system is more vulnerable to decoherence 
because the encoded logical state is an entangled state involving 
more physical qubits than the unprotected state, and thus is
exposed to more noise. 

\begin{figure}
   \includegraphics[width=0.48\columnwidth]{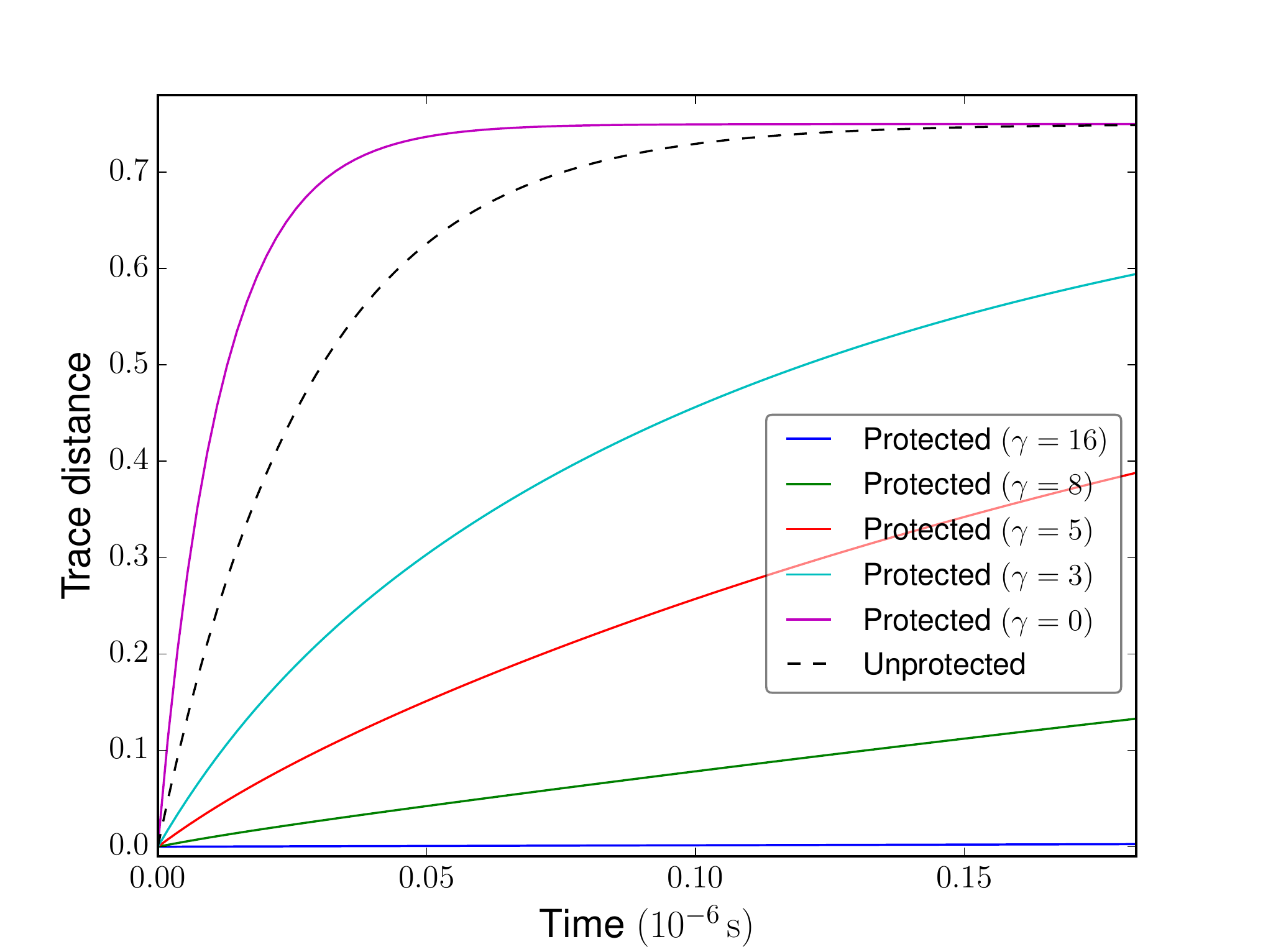}
   \includegraphics[width=0.48\columnwidth]{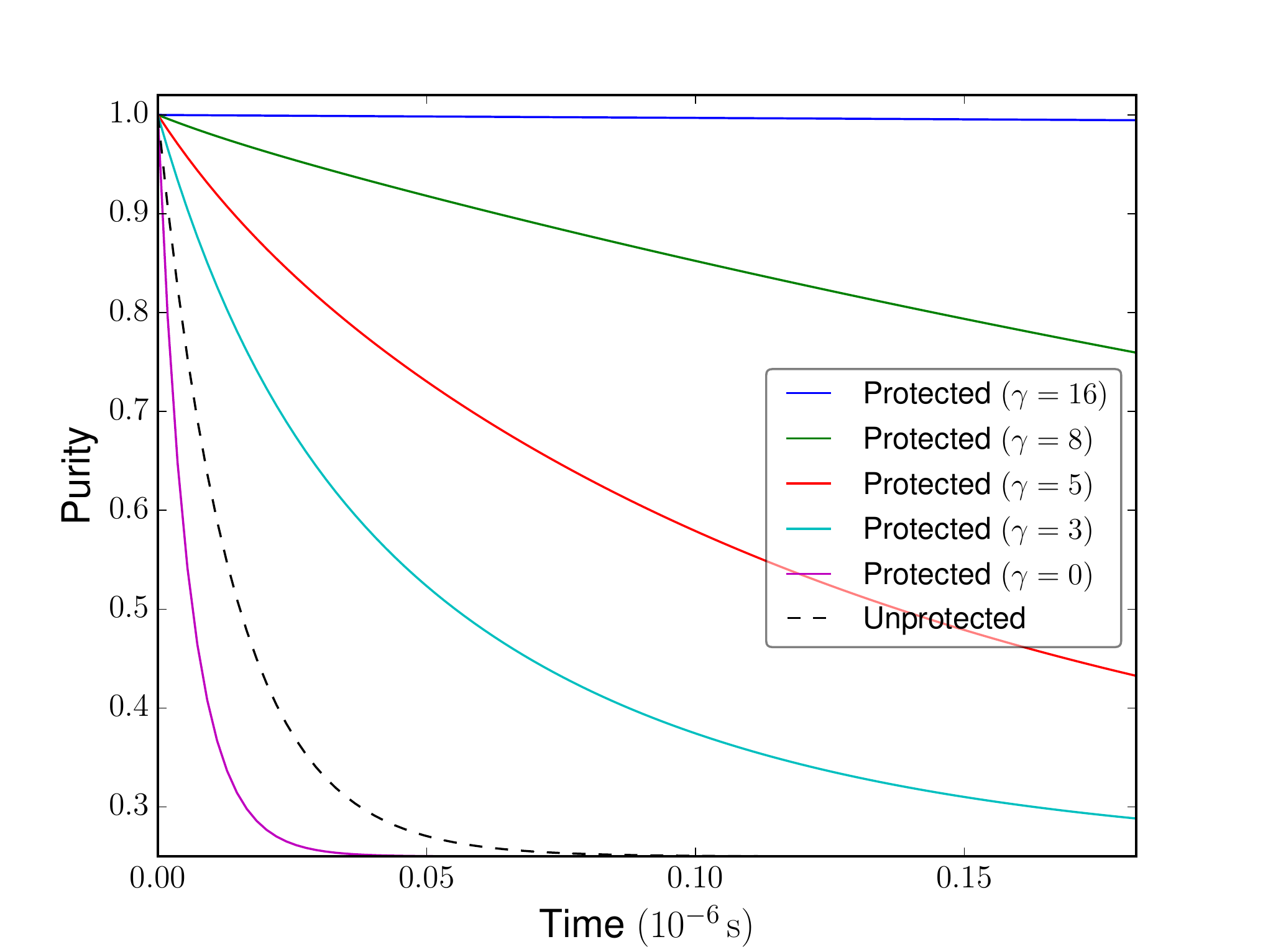}
   \caption{Simulations of the open-system dynamics of the two logical qubits,
encoded in the $[[6,2,2]]$ subsystem code, 
under single-qubit noise for different penalty weights $\gamma=\lambda/k_\mathrm{B}T$. 
The initial state of the logical
qubits is the Bell state $(\ket{00} + \ket{11})/\sqrt 2$.
(\textit{Left}) Trace distance between the initial and evolved state.  
(\textit{Right}) Purity of the evolved state.}
\label{fig:simulate_622}
\end{figure}

Figure~\ref{fig:simulate_622} shows, under the same noise model for various
values of $\gamma$, 
the purity of the state and the trace distance between 
the evolved and ideal states for two logical qubits, initially in a 
Bell state $(\ket{00} + \ket{11})/\sqrt 2$, encoded together in a single
block of the $[[6,2,2]]$ code. 

\begin{figure} 
   \includegraphics[width=0.48\columnwidth]{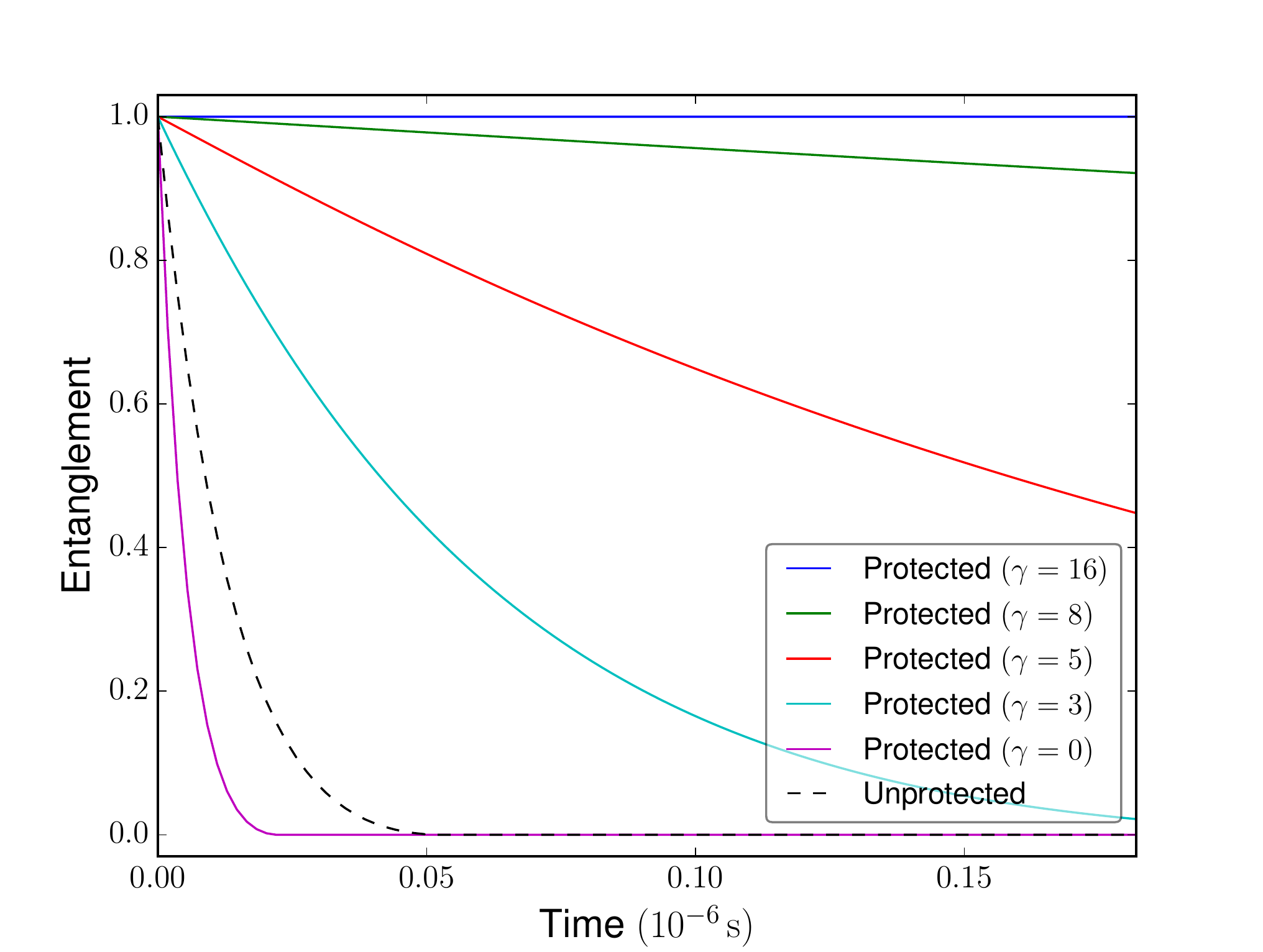}
   \includegraphics[width=0.48\columnwidth]{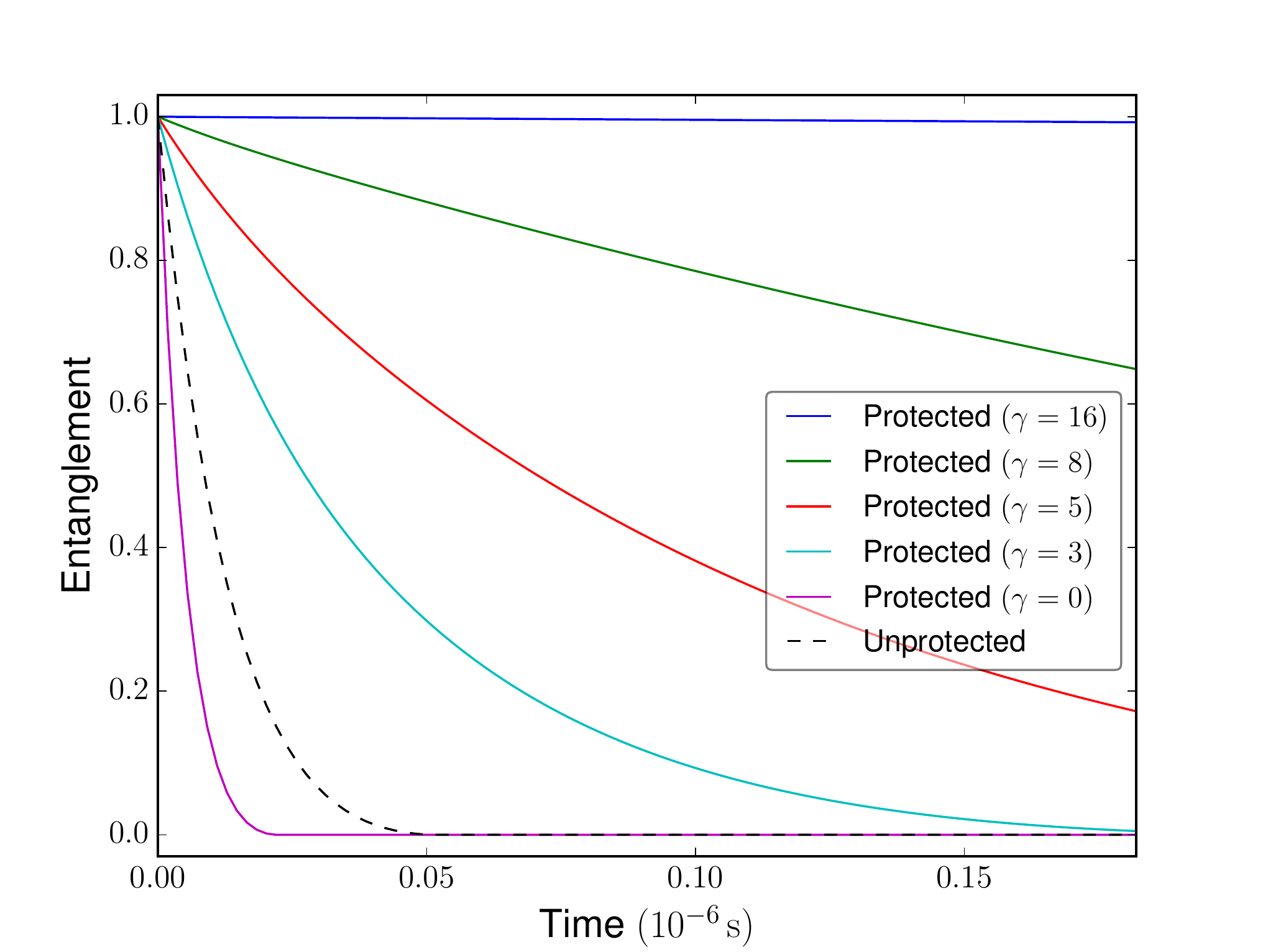}
   \caption{Simulations of the time evolution of entanglement between two
logical qubits. The initial state is set to be the Bell state $(\ket{00} +
\ket{11})/\sqrt 2$ of the logical qubits. (\textit{Left}) Two qubits encoded
separately using the $[[4,1,2]]$ code. (\textit{Right}) Two qubits encoded 
together using the $[[6,2,2]]$ code.}
\label{fig:entanglement}
\end{figure}

We also examined how well the two codes preserve entanglement. 
Figure~\ref{fig:entanglement} shows the entanglement of formation 
of two logical qubits, initially in a Bell state, each encoded
separately using (\textit{Left}) the $[[4,1,2]]$ code and (\textit{Right}) the $[[6,2,2]]$ code. 
Entanglement is better preserved by the $[[4,1,2]]$ code due to the
larger energy separation between its ground subspace and orthogonal subspaces.

\section{Conclusion} 

We have shown the effectiveness of a two-local Hamiltonian based
on subsystem codes in suppressing
single-qubit errors, thereby avoiding the no-go theorem that exists
in the stabilizer case.
We analyzed the energy separation between the ground subspace and 
orthogonal subspaces for some simple codes. As part of this analysis,
we developed a general technique to reduce
the dimension of the Hilbert space that needs to be considered, which eases
the calculation of the energy separation for arbitrary stabilizer
subsystem codes. We discussed the robustness of the subsystem-code-based
error suppression schemes with respect to implementation errors. 
We also numerically evaluated the effectiveness of these schemes in
suppressing errors coming from individual qubit interactions with
a spin-boson bath. 
One advantage of generalized-Bacon-Shor codes is the ability to 
implement certain two-qubit logical operators by involving only 
two physical qubits.  

We expect the subsystem error suppression approach discussed here 
to find applications in providing greater robustness for near-term 
quantum annealers, and also in the storage of quantum information, 
for example in 
quantum networks. More sophisticated techniques that would enable 
deeper analysis or more efficient numerical investigation of subsystem
code error suppression would be welcome. Of particular interest would
be techniques that enabled a scaling analysis of the energy separation
for families of subsystem codes, with the hopes of finding families of 
subsystem codes with asymptotically advantageous scaling of the
energy separation or ruling out such a possibility.

\begin{acknowledgements}
The authors would like to acknowledge support from the NASA Advanced 
Exploration Systems program and NASA Ames Research Center. 
This work was supported in part by 
the  AFRL Information Directorate under Grant F4HBKC4162G001,  
the Office of the Director of National Intelligence (ODNI), and 
the Intelligence Advanced Research Projects
Activity (IARPA), via IAA 145483.
The views and conclusions contained herein
are those of the authors and should not be interpreted as necessarily
representing the official policies or endorsements, either expressed
or implied, of ODNI, IARPA, AFRL, or the US Government.  The US
Government is authorized to reproduce and distribute reprints for
Governmental purpose notwithstanding any copyright annotation thereon.
\end{acknowledgements}



%

\appendix

\section{Rewriting the Hamiltonian in terms of auxiliary operators and
stabilizers}
\label{sec:auxiliary}

A critical step in calculating the energy separation for a 
two-local error-suppressing Hamiltonian arising from subsystem codes 
is rewriting the Hamiltonian in terms of auxiliary operators and stabilizers.  
This procedure makes explicit the dependence of the error-suppressing 
Hamiltonian on the values of the stabilizers. It also makes easier the 
analytical and numerical calculations of the energy separation by reducing
the size of the Hilbert space that needs to be considered 
and removes the degeneracy in the error-suppressing Hamiltonian. 
Another application of such a procedure enables exact diagonalization of a $6\times 6$ lattice of the quantum compass model~\cite{brzezicki_symmetry_2013}.

Here, we describe a systematic method for finding the 
$m-k$ $X$-type stabilizers, $m-k$ $Z$-type stabilizers, 
and $n - 2(m-k) - k$ $X$-type auxiliary logical operators,
and $n - 2(m-k) - k$ $Z$-type auxiliary logical operators,
defining $n - 2(m-k) - k$ auxiliary qubits 
for a $[[n,k,d]]$ generalized-Bacon-Shor code 
defined by a $m\times m$ binary matrix $M$ of rank $k$. 
We illustrate the application of this algorithm by using it to obtain
a set of auxiliary operators and stabilizers for the $[[16,2,3]]$ 
generalized-Bacon-Shor code.

We first present pseudocode for the algorithm and then comment on its workings.
\begin{algorithmic}
\State ${\cal A} \gets \emptyset$ \Comment Holds set of auxiliary operators
\State ${\cal S} \gets \emptyset$ \Comment Holds set of stabilizer generators

\Procedure{Row extraction}{$M$} 
\State Let the set ${\cal R}$ hold all the rows of $M$
\State ${\cal R}_{rem} \gets {\cal R}$ \Comment Holds remaining rows
\State ${\cal R}_{cur} \gets \emptyset$ \Comment Holds rows under consideration
\State Add the top row of $M$ to ${\cal R}_{cur}$
\While{${\cal R}_{rem}$ linearly dependent}
   \If{${\cal R}_{cur}$ linearly independent of 
${\cal R}_{rem}\setminus {\cal R}_{cur}$} 
      \State Move the rows in ${\cal R}_{cur}$ to the bottom of the matrix and 
        set ${\cal R}_{cur} \gets \emptyset$  
        \State Add the top row to ${\cal R}_{cur}$
   \EndIf 
    \State Add a minimally linearly independent set of rows in 
${\cal R}_{rem}\setminus {\cal R}_{cur}$ to ${\cal R}_{cur}$ to make the 
 top row linearly dependent on these rows and other rows in ${\cal R}_{cur}$
   \State Move these rows to the top 
of the matrix, above rows already in ${\cal R}_{cur}$
   \State Add to ${\cal S}$ the product of $Z$-type operators for every qubit 
in the rows of ${\cal R}_{cur}$
   \State Add $X$-type operators to ${\cal A}$ corresponding to all pairs 
of qubits in the top  row containing the first qubit 
   \State For all qubits in the top row except the first qubit, 
add $Z$-type operators to ${\cal A}$ that connect that
qubit with the next qubit below it in its column 
   \State Remove the top row from ${\cal R}_{cur}$ and ${\cal R}_{rem}$
\EndWhile
\State Form $M'$ from ${\cal R}_{rem}$
\EndProcedure

\Procedure{Column extraction}{$M'$} 
\State Let the set ${\cal C}'$ hold all the columns of $M'$
\State ${\cal C}_{rem} \gets {\cal C}'$  
\State ${\cal C}_{cur} \gets \emptyset$ 
\State Add the far left column of $M'$ to ${\cal C}_{cur}$
\While{${\cal C}_{rem}$ linearly dependent}
   \If{${\cal C}_{cur}$ linearly independent of ${\cal C}_{rem}\setminus {\cal C}_{cur}$}
      \State Move the columns in ${\cal C}_{cur}$ to the right side of the matrix and set \State ${\cal C}_{cur} \gets \emptyset$
      \State Add the far left column to ${\cal C}_{cur}$
   \EndIf
   \State Add a minimally linearly independent set of columns in ${\cal C}_{rem}\setminus {\cal C}_{cur}$ to ${\cal C}_{cur}$ to make it linearly dependent
   \State Move these columns to the left side of the matrix
   \State Add to ${\cal S}$ the product of $X$-type operators for every qubit 
in the corresponding columns to ${\cal C}_{cur}$ of the original matrix $M$ 
   \State Add $Z$-type operators to ${\cal A}$ corresponding to 
all pairs of qubits in the left-most column containing the first qubit 
   \For{each qubit in the left-most column except the first qubit}
   \State Let $A_X$ be the $X$-type operator connecting that qubit with the next qubit in its row 
   \For{$Z$-type operator $A'_Z$ in ${\cal A}$}
      \If{$A_X$ anti-commutes with $A'_Z$}
         \State Find the $X$-type operator $A'_X$ that anticommutes with $A'_Z$
         \State $A_X \gets A_X A'_X$
      \EndIf
   \EndFor 
   \EndFor
   \State Remove the left-most column from ${\cal C}_{cur}$ and 
${\cal C}_{rem}$
\EndWhile
\State Form $M''$ from ${\cal C}_{rem}$
\EndProcedure
\Procedure{Core extraction}{$M''$} 
\For{each pair of adjacent qubits in same column of $M''$}
   \State Let $A_Z$ be the $Z$-type operator for this pair
   \For{$X$-type operator $A'_X$ in ${\cal A}$}
      \If{$A_Z$ anti-commutes with $A'_X$}
         \State Find the $Z$-type operator $A'_Z$ that anticommutes with $A'_X$
         \State $A_Z \gets A_Z A'_Z$
      \EndIf
   \EndFor
   \State Add $A_Z$ to ${\cal A}$. 
\EndFor
\For{each adjacent pair of qubits in same row of $M''$}
   \State Let $A_X$ be the $X$-type operator for this pair
   \For{$Z$-type operator $A'_Z$ in ${\cal A}$}
      \If{$A_X$ anti-commutes with $A'_Z$}
         \State Find the $X$-type operator $A'_X$ that anticommutes with $A'_Z$
         \State $A_X \gets A_X A'_X$
      \EndIf
   \EndFor
   \State Add $A_X$ to ${\cal A}$
\EndFor
\EndProcedure
\end{algorithmic}

Switching the order of the rows of a matrix defining a generalized-Bacon-Shor
code does not change the gauge group, only the gauge generators.
Moving rows added to ${\cal R}_{cur}$ to the top of the matrix ensures
that the auxiliary $X$-type operators we define commute with all 
previously defined auxiliary operators. Specifically, the $Z$-type
operators previously defined do not involve this row.
The $Z$-type  operators we define are all within ${\cal R}_{cur}$, because 
the constraint of minimal linear dependence guarantees that there
will be a nonzero entry in the column below each qubit in the top row.
We exclude the left-most column $Z$-type operator because it is the product of the 
stabilizer and the other $Z$-type operators which we define. Because we
will not be including that column, we choose $X$-type operators
that include the first qubit since that enables us to satisfy easily the
canonical commutation relation for $X$- and $Z$-type operators defining
auxiliary qubits.  Once we are considering columns, we must be more careful. 
While the $Z$-type operators we define
automatically commute with all previously defined 
operators, because those rows were not used to extract $X$-type operators,
this property is not guaranteed for the $X$-type operators, which is why 
the extraction of $X$-type operators at this state is more complicated than 
for $Z$-type operators.

\subsection{Example: Extracting auxiliary operators and stabilizers for
the [[16,2,3]] code}
The following symmetric matrix defines a generalized-Bacon-Shor $[[16,2,3]]$ code,
\begin{align}
  M_{5\vert 5} = \kbordermatrix{
        & c_1 & c_2 & c_3 & c_4 & c_5 \\
    r_1 & 0   & 1   & 0   & 1   & 1 \\
    r_2 & 1   & 0   & 1   & 0   & 1 \\
    r_3 & 0   & 1   & 0   & 1   & 1\\
    r_4 & 1   & 0   & 1   & 0   & 1 \\
    r_5 & 1   & 1   & 1   & 1   & 0
  }\;.
\end{align}
The subscripts in $M_{5\vert 5}$ indicate the numbers of the rows and 
columns, respectively. We use $c_j$ and $r_k$ to label the row and 
column indices, so that we can keep track of the qubits after row and 
column manipulations.  The rank of the matrix $M_{5\vert 5}$ is $2$, 
and the distance of the code is $3$ (since any linear combination of 
its rows or columns has either at least three $1\,$s or none). We will
use the algorithm to find the $3$ $X$-type stabilizers,
the $3$ $Z$-type stabilizers, and the $8$ auxiliary qubits defined by
$8$ pairs of auxiliary operators satisfying the canonical commutation
relations.

\newcommand{\cur}[1]{\underline{#1}}
Moving row $r_3$ to the top of the matrix yields
\begin{align}
  M_{5\vert 5}' = \kbordermatrix{
          & c_1 & c_2 & c_3 & c_4 & c_5 \\
\cur{r_3} & 0   & 1   & 0   & 1   & 1 \\
\cur{r_1} & 0   & 1   & 0   & 1   & 1 \\
      r_2 & 1   & 0   & 1   & 0   & 1 \\
      r_4 & 1   & 0   & 1   & 0   & 1 \\
      r_5 & 1   & 1   & 1   & 1   & 0
  }\;,
\end{align} 
where we underline the row labels to indicate the rows (columns) we
are currently considering, rows that are in ${\cal R}_{cur}$.
Since the top two rows in $M_{5\vert 5}'$ are identical, and thus 
linearly dependent, we define a stabilizer
$S^Z_1 = R^Z_3  R^Z_1$, where $R^Z_3 = Z_{3,2}Z_{3,4}Z_{3,5}$ and $R^Z_1 = Z_{1,2}Z_{1,4}Z_{1,5}$.
We now extract auxiliary operators as 
we eliminate the top row. We define two auxiliary operators 
$X_{A1} = X_{3,2}X_{3,4}$ and $X_{A2} = X_{3,2}X_{3,5}$, and 
define the corresponding $Z$-type auxiliary operators to be 
$Z_{A1} = Z_{3,4}Z_{1,4}$ and $Z_{A2} = Z_{3,5}Z_{1,5}$.  
It is easy to check that two pairs of auxiliary operators satisfy 
the standard commutation relations.  
Note $Z_{3,2}Z_{1,2} = S^Z_1 Z_{A1} Z_{A2}$.  
Having used the top row to obtain auxiliary operators and stabilizers,
we may remove the top row. We consider the resulting matrix
\begin{align}
  M_{4\vert 5} = \kbordermatrix{
          & c_1 & c_2 & c_3 & c_4 & c_5 \\
\cur{r_1} & 0   & 1   & 0   & 1   & 1 \\
      r_2 & 1   & 0   & 1   & 0   & 1 \\
      r_4 & 1   & 0   & 1   & 0   & 1 \\
      r_5 & 1   & 1   & 1   & 1   & 0
  }\;.
\end{align}
A minimally linearly independent set on which the top row is linearly
dependent is $\{r_2, r_5\}$. We move these rows to the top of the
matrix to obtain
\begin{align}
  M'_{4\vert 5} = \kbordermatrix{
          & c_1 & c_2 & c_3 & c_4 & c_5 \\
\cur{r_2} & 1   & 0   & 1   & 0   & 1 \\
\cur{r_5} & 1   & 1   & 1   & 1   & 0 \\
\cur{r_1} & 0   & 1   & 0   & 1   & 1 \\
      r_4 & 1   & 0   & 1   & 0   & 1 
  }\;.
\end{align}
We define a stabilizer $S^Z_2 = R^Z_2 R^Z_5 R^Z_1$, where 
$R^Z_2 = Z_{2,1}Z_{2,3}Z_{2,5}$, $R^Z_5 =  Z_{5,1}Z_{5,2}Z_{5,3}Z_{5,4}$, 
and $R^Z_1 =Z_{1,2}Z_{1,4}Z_{1,5}$.
We define $X$-type  auxiliary operators
$X_{A3} = X_{2,1}X_{2,3}$ and $X_{A4} = X_{2,1}X_{2,5}$, and $Z$-type
auxiliary operators
$Z_{A3} = Z_{2,3}Z_{5,3}$, and $Z_{A4} = Z_{2,5}Z_{1,5}$.  
We may now remove the top row to obtain
\begin{align}
  M_{3\vert 5} = \kbordermatrix{
          & c_1 & c_2 & c_3 & c_4 & c_5 \\
\cur{r_5} & 1   & 1   & 1   & 1   & 0 \\
\cur{r_1} & 0   & 1   & 0   & 1   & 1 \\
      r_4 & 1   & 0   & 1   & 0   & 1 
  }\;.
\end{align}
We move $r_4$ to the top of the
matrix to obtain
\begin{align}
  M_{3\vert 5}' = \kbordermatrix{
          & c_1 & c_2 & c_3 & c_4 & c_5 \\
\cur{r_4} & 1   & 0   & 1   & 0   & 1 \\    
\cur{r_5} & 1   & 1   & 1   & 1   & 0 \\
\cur{r_1} & 0   & 1   & 0   & 1   & 1      
  }\;.
\end{align}
We define a stabilizer $S^Z_3 = R^Z_4 R^Z_5 R^Z_1 $, where $R^Z_4 =Z_{4,1}Z_{4,3}Z_{4,5}$,
$R^Z_5 =  Z_{5,1}Z_{5,2}Z_{5,3}Z_{5,4}$, and $R^Z_1 = Z_{1,2}Z_{1,4}Z_{1,5}$. 
We define $X$-type auxiliary operators $X_{A5} = X_{4,1}X_{4,3}$ and
$X_{A6} = X_{4,1}X_{4,5}$, and $Z$-type operators $Z_{A5} = Z_{4,3}Z_{5,3}$ and $Z_{A6} = Z_{4,5}Z_{1,5}$. We may now remove the top row to obtain 
\begin{align}
  M_{2\vert 5} = \kbordermatrix{
        & c_1 & c_2 & c_3 & c_4 & c_5 \\
    r_5 & 1   & 1   & 1   & 1   & 0 \\    
    r_1 & 0   & 1   & 0   & 1   & 1 
  }\;.
\end{align}
Now that the rows are linearly independent, we can engage ``column elimination'' 
to extract further operators.  We move $c_3$ to the far left to obtain
\begin{align}
  M_{2\vert 5}' = \kbordermatrix{
        & \cur{c_3} & \cur{c_1}    & c_2  & c_4 & c_5 \\
    r_5 & 1         & 1   & 1      & 1   & 0 \\    
    r_1 & 0         & 0   & 1      & 1   & 1 
  }\;.
\end{align}
Were $M_{2\vert 5}'$ the starting matrix, the first $X$-type stabilizer 
would be $\tilde S^X_1 = X_{5,3}X_{5,1}$. 
However, this $X$-type stabilizer does not commute with some of 
the $Z$-type auxiliary operators we introduced before.  
To obtain the correct stabilizer, we need to iteratively
multiply $\tilde S^X_1$ by $X$-type operators corresponding to 
(anti-commuting with) $Z$-type operators that do not commute with 
$\tilde S^X_1$. It does not commute with $Z_{A5} = Z_{4,3}Z_{5,3}$, so we 
need to multiply by $X_{A5} = X_{4,1}X_{4,3}$, which in turn does not
commute with $Z_{A3} = Z_{2,3}Z_{5,3}$, so we need to multiply by
$X_{A3} = X_{2,1}X_{2,3}$. The result is the stabilizer
\begin{align}
\begin{split}
 S^X_1 
 &= X_{2,3}X_{4,3} X_{5,3} X_{2,1} X_{4,1} X_{5,1} \\
 &= C^X_3 C^X_1\;.
 \end{split}
\end{align}
The left-most column contains
no pairs of $1\,$s, so we do not extract any auxiliary operators at this step.
We may now remove the far left column to obtain
\begin{align}
  M_{2\vert 4} = \kbordermatrix{
         & \cur{c_1}    & c_2  & c_4 & c_5 \\
    r_5  & 1   & 1      & 1   & 0 \\    
    r_1  & 0   & 1      & 1   & 1 
  }\;.
\end{align}
Columns $c_2$ and $c_5$ form a minimally linearly independent set on which
the left-most column $c_1$ depends. We therefore move these columns to
the far left to obtain
\begin{align}
  M_{2\vert 4}' = \kbordermatrix{
         & \cur{c_2}  & \cur{c_5}  & \cur{c_1} & c_4 \\
    r_5  & 1   & 0    & 1   & 1 \\    
    r_1  & 1   & 1    & 0   & 1 
  }\;.
\end{align}
A similar argument to the one above leads to defining the stabilizer
\begin{align}
\begin{split}
 S^X_2 
 &= X_{1,2} X_{3,2} X_{5,2} X_{1,5} X_{2,5} X_{3,5} X_{4,5} X_{2,1} X_{4,1} X_{5,1}\\
 &= C^X_2 C^X_5 C^X_1 \;.
\end{split}
\end{align}
We also define the pair of operators $Z_{A7}=Z_{5,2}Z_{1,2}$ and $\tilde X_{A7}=X_{1,2}X_{1,5}$. Since $\tilde X_{A7}$ anticommutes with
$Z_{A2}$, $Z_{A4}$, and $Z_{A6}$, we have 
$X_{A7}=\tilde X_{A7}X_{A2}X_{A4}X_{A6}$.
We now remove the left-most column to obtain
\begin{align}
  M_{2\vert 3} = \kbordermatrix{
         & \cur{c_5}  & \cur{c_1} & c_4 \\
    r_5  & 0    & 1   & 1 \\    
    r_1  & 1    & 0   & 1 
  }\;.
\end{align}
Moving the row $c_4$ to far left, we have
\begin{align}
  M_{2\vert 3}' = \kbordermatrix{
         & \cur{c_4} & \cur{c_5} & \cur{c_1}  \\
    r_5  & 1         & 0         &1    \\    
    r_1  & 1         & 1         & 0   
  }\;.
\end{align}
These columns contribute a stabilizer
\begin{align}
\begin{split}
 S^X_3 &= C^X_4 C^X_5 C^X_1  \\
 &= X_{1,4} X_{3,4}X_{5,4} X_{1,5}X_{2,5}X_{3,5}X_{4,5} X_{2,1}X_{4,1} X_{5,1}\;.
\end{split}
\end{align}
We also extract the final set of auxiliary operators
$Z_{A8} = Z_{5,4}Z_{1,4}$ and $X_{A8} = \tilde X_{A8} X_{A1} X_{A2} X_{A4}X_{A6}$, where $\tilde X_{A8} = X_{1,4}X_{1,5}$. 
Removing the first column results in a matrix with linearly independent
rows and columns and in which no row or column contains a pair of $1\,$s; 
therefore we stop extracting operators. Indeed, we have already 
obtained $3$ $X$-type and $3$ $Z$-type stabilizers and $8$ pairs of
auxiliary operators as expected. 

In conclusion, we have following auxiliary operators
\beq
\begin{gathered}
 X_{A1} = X_{3,2}X_{3,4}\,,\quad X_{A2} = X_{3,2}X_{3,5}\,,\quad X_{A3} = X_{2,1}X_{2,3}\,,\\
 X_{A4} = X_{2,1}X_{2,5}\,,\quad X_{A5} = X_{4,1}X_{4,3}\,,\quad Z_{A1} = Z_{1,4}Z_{3,4}\,,\\
 Z_{A2} = Z_{1,5}Z_{3,5}\,,\quad Z_{A3} = Z_{2,3}Z_{5,3}\,,\quad Z_{A4} = Z_{2,5}Z_{1,5}\,,\\
 Z_{A5} = Z_{4,3}Z_{5,3},\,\quad X_{A6} = X_{4,1}X_{4,5}\,,\quad 
 X_{A7}=X_{1,2}X_{1,5}X_{A2}X_{A4}X_{A6}\,,\\\quad X_{A8} = X_{1,4}X_{1,5}X_{A1} X_{A2} X_{A4} X_{A6} \,,\quad  Z_{A6} = Z_{4,5}Z_{1,5}\,,\\
 Z_{A7}=Z_{5,2}Z_{1,2}\,,\quad Z_{A8} = Z_{5,4}Z_{1,4}\,,
\end{gathered}
\eeq
and the stabilizers take the form
\beq
\begin{gathered}
 S^X_1 = C^X_3C^X_1\,,\quad S^X_2 = C^X_2C^X_5C^X_1\,,\quad   S^X_3 =  C^X_5 C^X_1 C^X_4\,,\\
 S^Z_1 = R^Z_1 R^Z_3\,,\quad S^Z_2 = R^Z_2R^Z_5 R^Z_1\,,\quad  S^Z_3 = R^Z_5 R^Z_1 R^Z_4\,.
\end{gathered}
\eeq

\section{Noise model}
\label{sec:noise_model}
For our numerical analyses of error suppression,
we consider the spin-boson Hamiltonian~\cite{leggett_dynamics_1987},
\begin{align}\label{eq:total_hamiltonian}
 H(t) =  H_S(t) + H_B + \sum_{k=1}^n \Big(X_k\otimes B_k^X+  Y_k\otimes B_k^Y + Z_k\otimes B_k^Z\Big)\;,
\end{align}
where $H_S = H_\supp + H_L$ is the system Hamiltonian, and $X_k$, 
$Y_k$, and $Z_k$ are Pauli operators acting on the $k$th qubit. 
The sum in Eq.~(\ref{eq:total_hamiltonian}) describes interactions between
individual Pauli operators of the system qubits and independent bath modes,
where \begin{gather}
 B_k^X = \sum_\mu g_\mu^X\big( \a_{\mu, k} + \a_{\mu, k}^\dagger \big)\;,\\
 B_k^Y = \sum_{\nu} g_\nu^Y  \big(\b_{\nu, k} + \b_{\nu, k}^\dagger \big)\;,\\
 B_k^Z = \sum_{\tau} g_\tau^Z  \big(\c_{\tau, k} + \c_{\tau, k}^\dagger \big)\;,
\end{gather}
with $g_\mu^X$, $g_\nu^Y$, and $g_\tau^Z$ being the coupling constants. 
We consider the case in which
all of these coupling constants have the same value.
The term $H_B$ in Eq.~(\ref{eq:total_hamiltonian}) is the bath Hamiltonian,
\begin{align}
 H_B = \sum_{\mu, j} \hbar\, \omega_{\mu, j}^X\, \a^\dagger_{\mu, j} \a_{\mu, j}  + \sum_{\nu, k} \hbar \, \omega_{\nu, k}^Y\, \b^\dagger_{\nu, k} \b_{\nu, k}  + \sum_{\tau, l}\hbar\,  \omega_{\tau, l}^Z\, \c^\dagger_{\tau, l} \c_{\tau, l}\;.
\end{align}
Going to the Heisenberg picture of the bath Hamiltonian, we have
\begin{align}
 B_k(t) = e^{itH_B/\hbar} B_k\, e^{-itH_B/\hbar} = \sum_\mu \Big(  g_\mu\a_{\mu, k} e^{-i\omega_\mu t} +  g_\mu^*\a_{\mu, k}^\dagger e^{i\omega_\mu t}\Big)\;,
\end{align}
where the superscripts $X$, $Y$, and $Z$ are neglected for abbreviation of 
notation.  The bath correlation function then takes the form
\begin{align}\label{eq:correlation}
\begin{split}
C_\mathrm{bath}(j,t;\,k,t') &=\big\langle B_j(t)  B_k(t')\big\rangle\\
&=  \delta_{j,k}\sum_\mu\, \norm{g_\mu}^2 \Big(  \big\langle\a_{\mu, k}^\dagger \a_{\mu, k}\big\rangle\, e^{-i\omega_{\mu, k} (t-t')} +  \mathrm{c.c.}\Big)\\
&= \delta_{j,k} C_\mathrm{bath}(t-t')\;.
\end{split}
\end{align}
The expectation values in Eq.~(\ref{eq:correlation}) satisfy the Planck condition for thermal baths,
\begin{align}
 &\big\langle\a_{\mu}^\dagger\ssp \a_{\mu}\big\rangle = \frac{1}{e^{\hbar\omega_\mu/k_\mathrm{B} T}-1}\;, \quad\; \big\langle\a_{\mu}\ssp \a_{\mu}^\dagger\big\rangle = \big\langle\a_{\mu}^\dagger\ssp \a_{\mu}\big\rangle + 1 = \frac{1}{1-e^{-\hbar\omega_\mu/k_\mathrm{B} T}}\;,
\end{align}
where the qubit subscript $k$ is omitted. 
The Fourier transformation of the bath correlation function is
\begin{align}
 \widetilde C_\mathrm{bath}(\omega)  =  \int \dif t\, e^{-i\omega t}C_\mathrm{bath}(t)= \frac{2 \pi J(\norm{\omega})}{\left\vert 1- e^{-\hbar\omega/k_\mathrm{B} T}\right\vert}\;, 
\end{align}
where $J(\omega)$ is the bath spectral function arising from the 
substitution of the sum in Eq.~(\ref{eq:correlation}) with an integral,
\begin{align}
 \sum_\mu \norm{g_\mu}^2 \simeq \int_0^\infty \dif \omega\, J(\omega)\;.
\end{align}
The bath correlation function determines the transition rate from one system state $\ket{\psi_\alpha}$ to another state $\ket{\psi_\beta}$,
where $E_\alpha$ and $E_\beta$ are the energies of the two states and $\noise$ is the system noise operator (in this case $X$, $Y$, and $Z$).  The ratio of the transition rates between any two states satisfies
\begin{align}
\frac{\Gamma_{\psi_\alpha \rightarrow
\psi_\beta}}{
\Gamma_{\psi_\beta \rightarrow 
\psi_\alpha}} = e^{(E_\alpha-E_\beta)/k_\mathrm{B} T}\;,
\end{align}
which gives the correct population ratio of $\ket{\psi_\beta}\bra{\psi_\beta}$ and $\ket{\psi_\alpha}\bra{\psi_\alpha}$ at thermal equilibrium, i.e., the Boltzmann distribution. The function $\widetilde C_\mathrm{bath}(\omega)$ determines the transition rate from a lower-energy state to a higher-energy state when $\omega <0$, and the other way around when $\omega > 0$.  While transitions to higher-energy states are detrimental, 
transitions to lower-energy states are beneficial for adiabatic
quantum computation.

We further assume that the bath spectral function satisfies the Ohmic condition,
\begin{align}
 J (\omega) \simeq  \hbar^2 \chi\, \omega\ssp e^{-\omega/\omega_c}\,,\quad \mathrm{for}\;\omega \geq 0\;,
\end{align}
where $\chi$ is a dimensionless constant and $\omega_c$ is the cutoff frequency. The bath correlation function can thus be simplified to 
\begin{align}\label{eq:ohmic_noise}
 \widetilde C_\mathrm{bath}(\omega)  =  \frac{2 \pi \hbar^2 \chi \omega\ssp e^{-\norm{\omega}/\omega_c}}{1- e^{-\omega/\omegaT }}\;,
\end{align}
where $\omegaT = k_\mathrm{B}T/\hbar$.  We plot this function with the parameters given in~\cite{pudenz_error-corrected_2014}: $\chi = 3.18 \times 10^{-4}$, $\omega_c=8\pi\times 10^9\, \mathrm{rad/s}$, and $\omegaT = 2.2\times 10^9\, \mathrm{rad/s}$ (at $17\,\mathrm{mK}$). 
\begin{figure}[ht] 
   \centering
   \includegraphics[width=0.7\linewidth,keepaspectratio]{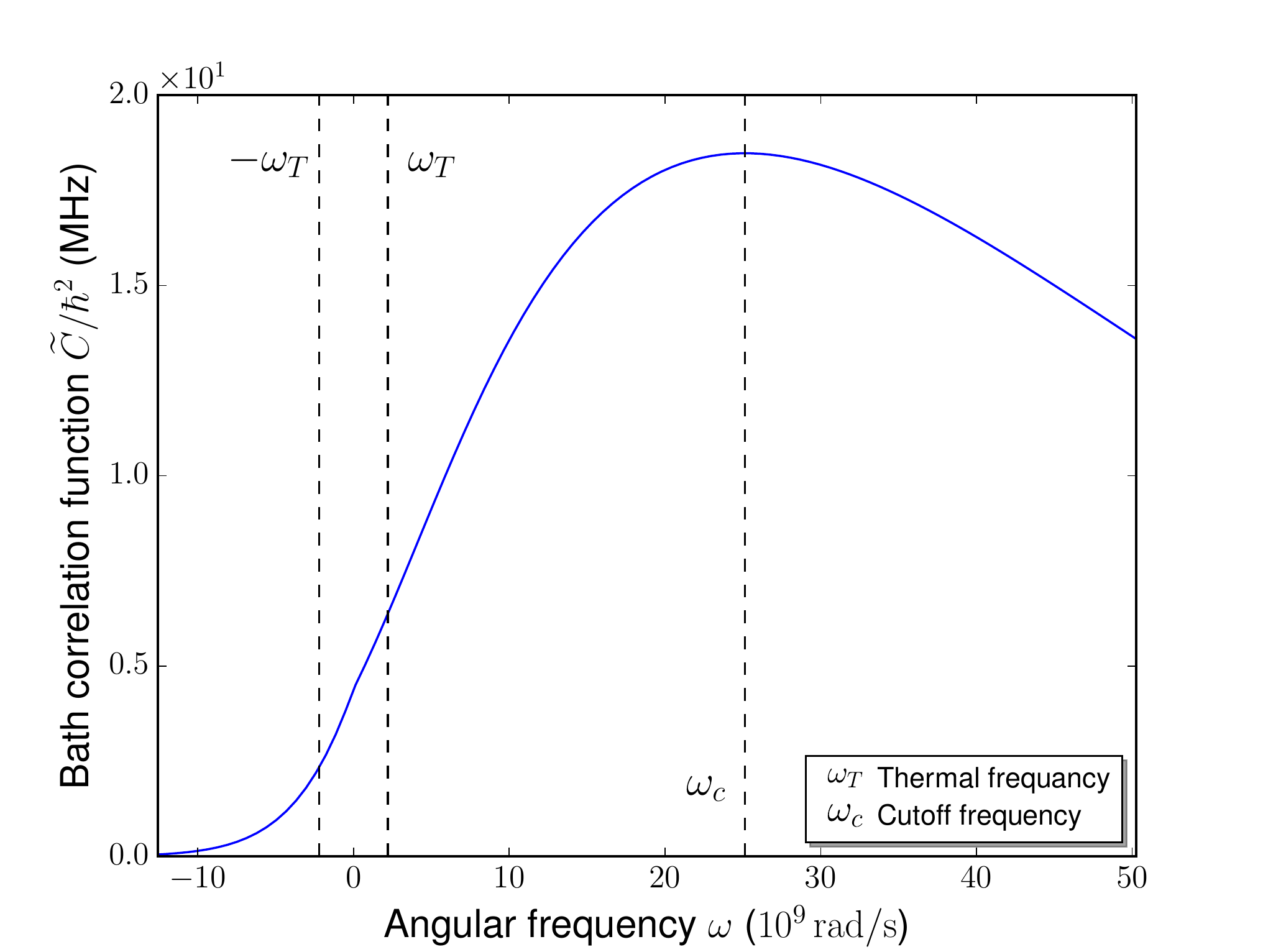}
   \caption{Bath correlation function for Ohmic noise}
   \label{fig:bath_corr_fun}
\end{figure}
At zero frequency, i.e., transitions between two states of the same energy, $\widetilde C_\mathrm{bath}(0) = 2 \pi \hbar^2 \chi\, \omegaT$ is proportional to the temperature $T$. The derivative of $\widetilde C_\mathrm{bath}(\omega)$ is not continuous at $\omega = 0$ due to the finite cutoff frequency $\omega_c$. 
The function $\widetilde C_\mathrm{bath}(\omega)$ decays 
quickly once $\omega$ is
smaller than $-\omegaT$; the transition rate to higher-energy states 
is low for an energy difference that is several times larger than $k_\mathrm{B}T$.
Consequently, an energy gap as large as several times of $k_\mathrm{B} T$ can keep the system in the ground state for a much longer time than the gapless case. 
The asymmetry of the function $\widetilde C_\mathrm{bath}(\omega)$ 
(see Figure~\ref{fig:bath_corr_fun})
can also be used to prepare the initial state when the energy gap is less 
than $\hbar\omega_c$ but larger than $\hbar\omegaT$, as the noise terms 
drive the system to its ground state while the opposite effect is suppressed.

\end{document}